# Numerical Study on the Performance of Perforated Breakwater for Green Water

**Ph.D. Student Rubens Augusto Amaro Junior (Corresponding author)**

Department of Construction Engineering, Polytechnic School of the University of São Paulo

Av. Prof. Almeida Prado, trav. 2, 83 - Cidade Universitária, 05508-070, São Paulo, SP, Brazil

e-mail: rubens.amaro@usp.br

**Professor Liang-Yee Cheng**

Department of Construction Engineering, Polytechnic School of the University of São Paulo

Av. Prof. Almeida Prado, trav. 2, 83 - Cidade Universitária, 05508-070, São Paulo, SP, Brazil

e-mail: cheng.yee@usp.br

**Graduate Student Stella Vieira Rosa**

Department of Mechanical Engineering, Polytechnic School of the University of São Paulo

Av. Professor Melo Moraes, 2231 - Cidade Universitária, 05508-030, São Paulo, SP, Brazil

e-mail: stella.rosa@usp.br

**Abstract.** In this work, the influence of the geometry of lightweight perforated breakwaters on their performance against green water impact loads is investigated systematically through a series of numerical simulations. For this purpose, a fully Lagrangian meshfree particle-based method is adopted to model transient and impulsive hydrodynamic phenomenon with free surface and complex geometry. The green water flow is represented by a dam-break problem; the breakwater is modeled as a perforated plate and the protected installation as a vertical wall. Computed impact force, moment, and impulse on the breakwater and protected wall show nonlinear effects, and the main geometric parameter is the open-area ratio. In addition, the increase in the breakwater height is effective only for low open-area ratios. The influence of the arrangement of the holes are also investigated. Different flow patterns are obtained for small and large gaps between the breakwater and the wall. The relation between the loads on the breakwater and the wall also provides the basis for the optimal design of the breakwater, considering the tolerant hydrodynamic loads of protected structures.

**Introduction**

Environmental conditions, such as wind, current and waves are of great concern in the design of coastal and offshore structures due to safety and operational issues. Among the environmental factors, hydrodynamic loads due to the waves are the most critical ones. In this way, wave breakers or breakwaters have been used worldwide as protection devices to mitigate hydrodynamic loads.

Nowadays, the optimal design of the breakwaters by considering the wave conditions and the tolerable loads on the protected installations or equipment of the coastal and offshore structures still remains as a challenge. This is more relevant in case of the green water protection for floating structures, for example floating liquefied natural gas units (FLNG) or very large floating structures (VLFS) such as a floating airport, where the weight of the protection devices needs to be considered in the design.

As the weight of the protection device is generally restricted, lightweight protection devices onboard the floating structures, such as vane type devices and perforated plates have been proposed in the past. However, there are relatively few literatures regarding the systematic study on the performance of these devices as well as the influences of their geometrical parameters to mitigate the loads due green water. Among the works, we have: The investigation of Fekken (1998) on the experimental and numerical pressures and forces on different structural shapes; Buchner (2002) adopted a semi-empirical design evaluation method to investigate the run-up water height and loads on the protection device; Varyani et al. (2005) and Pham and Varyani (2006) investigated numerically the variation of geometrical parameters of the breakwaters; Silva and Rossi (2014) regarding numerical simulations of vane type devices with some variations in the geometric parameters and the number of vanes. A reason for this restricted study on the subject is the complicated experimental settings required and the limitations of conventional mesh-based Computational Fluid Dynamics (CFD) approaches, such as Finite Volume Method (FVM), Finite Element Method (FEM) and Boundary Element Method (BEM), for the modeling of the highly nonlinear hydrodynamic phenomena that involve large free-surface deformation, wave breaking, flow separation and coalescence and relatively complex geometry of the protection devices.

Recently, due to remarkable advances on the high-performance computing (HPC) systems, several research have shown that the flexible but compute-intensive, particle-based methods are promising approaches for the investigation of green water. Within this context, the macroscopic scale particle-based modeling that solves the governing equation of continuum, such as Smoothed Particle Hydrodynamics (SPH) method (Gingold and Monagham 1977; Lucy 1977) and Moving Particle Semi-implicit (MPS) method (Koshizuka and Oka 1996) are performing relevant roles. As meshless method, they are very effective for the simulation of problems involving complex deformations of the boundaries. To mention few regarding SPH method, we have: Gómez-Gesteira et al. (2005) modeled the impact of a single wave on a flat horizontal deck to study the green water overtopping, of which the computed wave phase and amplitude showed good agreement with experimental results; Pakozdi et al. (2012) used inlet condition to model water propagation on deck of floating production storage and offloading unit (FPSO) in irregular waves and water kinematics similar to results based on Volume of Fluid (VOF) methods were obtained, as well as reasonable agreement with the experimental ones. In Guilcher et al. (2013), SPH method was coupled to potential-theory to simulate green water on a simplified fixed FLNG. Despite underestimation of water heights and pressures at some probes, good agreement with experimental values was achieved. About the investigations based on MPS method, Shibata et al. (2012) successfully reproduced the nonlinear effect of shipping water. Nevertheless, due spatial resolution and tank dimensions, quantitative differences between the computed and experimental results still remained. Bellezi et al. (2013) investigated numerically the effects of the bow shape on the green water loads with good agreement with the experimental results. Zhang et al. (2013) analyzed the evolution of wave shipping on the deck. The computed impact pressure on the deck structures also agreed well with experimental data and other numerical results.

In the present work, the influence of the geometrical parameters of the perforated breakwaters, such as shape, size, quantity and arrangement of the holes, the height of the breakwater and its distance to the protected device are investigated systematically through a series of numerical simulations. The relations between the loads on the protection device and the protected ones are also shown to provide insights for the optimal design of the perforated lightweight breakwaters subjected to green water impact loads. For this purpose, a fully Lagrangian meshfree particle-based approach based on MPS method is adopted.

Focusing on the investigation of the basic issues of the localized nonlinear hydrodynamic phenomenon, the green water flow on deck is represented herein by a dam breaking problem in the present study. This was done, for example, by Goda and Miyamoto (1976), Dillingham (1981), Mizoguchi (1988), Buchner (2002), Nielsen and Mayer (2004), and Pham and Varyani (2005) to determine the amount of water, pressures or forces on the deck, neglecting the ship motion. The lightweight breakwaters considered here are perforated plates. Also, for sake of simplicity, the protected installation is modeled as a vertical wall. First, convergence analysis is carried out to optimize the numerical modeling. After that, hydrodynamic loads on the breakwater and protected wall are obtained and compared for different geometrical configurations of the breakwater and possible correlations are discussed. Moreover, the results also provide elements for the optimal design of breakwater considering the tolerant hydrodynamic impact loads of the protected structures.

**Numerical Method**

The governing equations of incompressible viscous flow are expressed by the conservation laws of mass and momentum:

$$\frac{D\rho}{Dt} = -\rho \nabla \cdot \mathbf{u} = 0, \tag{1}$$

$$\frac{D\mathbf{u}}{Dt} = -\frac{\nabla P}{\rho} + \nu \nabla^2 \mathbf{u} + \mathbf{f}, \tag{2}$$

where $\rho$ is the fluid density, $\mathbf{u}$ is the velocity vector, $P$ is the pressure, $\nu$ is the kinematic viscosity and $\mathbf{f}$ is the external body force per unit mass vector.

*Discrete differential operators*

In MPS method, in which all the computational domain, including solid and fluid, is discretized in Lagrangian particles, the differential operators of the governing equations are replaced by discrete differential operators on irregular nodes (Isshiki 2011), which are derived based on a weight function. For a given particle $i$, the influence of a neighbor particle $j$ is defined by weight function $\omega_{ij}$:

$$\omega_{ij} = \begin{cases} \dfrac{r_e}{\|\mathbf{r}_{ij}\|} - 1 & \|\mathbf{r}_{ij}\| \le r_e \\ 0 & \|\mathbf{r}_{ij}\| > r_e \end{cases}, \tag{3}$$

where $r_e$ is the effective radius that limits the range of influence and $\|\mathbf{r}_{ij}\|$ is the distance between $i$ and $j$. For the support of the weight function, as demonstrated by Koshizuka and Oka (1996), more accurate and stable computations can be achieved with $r_e \ge 1.8 l_0$, where $l_0$ is the initial distance between two adjacent particles. Concerning circular support of 2D problems, the authors showed that the effective radius for lower order operators, such as gradient or calculation of particle number density, should be $r_e < 3.0 l_0$, while for second order operators, such as Laplacian, requires larger support, normally $r_e \ge 3.1 l_0$. However, in 3D problems the spherical support leads to larger number of neighboring particles and provides good numerical approximation for a smaller $r_e$. Therefore, in the present work, $r_e = 2.1 l_0$ is adopted for the calculations of particle number density and all differential operators.

The summation of the weight of all the particles in the neighborhood of the particle $i$ is defined as its particle number density $n_i$, which is proportional to the fluid density:

$$n_i = \sum_{j \ne i} \omega_{ij}. \tag{4}$$

For a scalar function $\phi$, the gradient and Laplacian operators are defined in Eqs. (5) and (6), respectively:

$$\nabla \phi = \frac{d}{n^0} \sum_{j \ne i} \frac{\phi_j - \phi_i}{\|\mathbf{r}_{ij}\|^2} \mathbf{r}_{ij} \omega_{ij}, \tag{5}$$

$$\nabla^2 \phi = \frac{2d}{\lambda_i n^0} \sum_{j \ne i} (\phi_j - \phi_i) \omega_{ij}, \tag{6}$$

where $d$ is the number of spatial dimensions and $n^0$ is the initial value of particle number density for a complete support of neighbor particles. Finally, $\lambda_i$ is a correction parameter so that the variance increase is equal to that of the analytical solution, and is calculated by:

$$\lambda_i = \frac{\sum_{j \ne i} \omega_{ij} \|\mathbf{r}_{ij}\|^2}{\sum_{j \ne i} \omega_{ij}}. \tag{7}$$

*Algorithm*

To solve the incompressible viscous flow, a semi-implicit algorithm similar to the projection method (Chorin 1967) is adopted in the MPS method by using Euler explicit time integration of velocity and implicit calculation of pressure. First, explicit predictions of the particle's velocity and position are carried out by using viscosity and external forces in terms of the momentum conservation, Eq. (2). Then the pressure of all particles is calculated by the pressure Poisson equation (PPE) as follows:

$$\nabla^2 P_i^{t+\Delta t} - \frac{\rho}{\Delta t^2} \alpha P_i^{t+\Delta t} = -\gamma \frac{\rho}{\Delta t^2} \frac{n_i^* - n^0}{n^0}, \qquad (8)$$

where $n_i^*$ is the particle number density calculated based on the displacement of particles obtained in the prediction step, $\alpha$ is the coefficient of artificial compressibility and $\gamma$ is the relaxation coefficient. Both $\alpha$ and $\gamma$ are used to improve the stability of a computation method. Also, to provide more stable results, the pressure gradient is modified as (Koshizuka and Oka 1996):

$$\nabla P = \frac{d}{n^0} \sum_{j \neq i} \frac{P_j - \hat{P}_i}{\|\mathbf{r}_{ij}\|^2} \mathbf{r}_{ij} \omega_{ij}, \qquad (9)$$

where $\hat{P}_i$ is the minimum pressure between the neighborhood of the particle $i$.

Finally, the velocity of the particles is updated by using the pressure gradient term of the momentum conservation and the new positions of the particles are obtained.

*Fixed rigid wall*

Since the focus of this work is to study the influences of the geometric parameters of perforated breakwater, displacement and deformation of the structure were not taken into account. In this way, the coupling between fluid and the rigid and fixed structure is done directly by the application of the no-slip condition on the wall, which is considered in the calculation of pressure and flow velocity.

Fixed rigid wall boundary is modeled by using three layers of fixed particles. The particles that form the layer in contact with the fluid are denominated wall particles, of which the pressure is computed by solving PPE, see Eq. (8), together with the fluid particles. The particles that form two other layers are denominated dummy particles.

Dummy particles are used to assure the correct calculation of the particle number density of the wall particles. Pressure is not calculated in the dummy particles.

*Free surface*

In the Lagrangian particle-based method, the kinematic boundary condition of a free surface is directly satisfied by motion of the free surface particles. On the other hand, the dynamic boundary condition is imposed by applying the Dirichlet condition:

$$P_{fs} = P_{atm} = 0, \tag{10}$$

where $P_{fs}$ is the pressure at free surface particles and $P_{atm}$ stands the atmospheric pressure.

In order to identify free surface particles, the particle number density is used as a checking parameter. A particle is defined as free surface particle and its pressure is set to zero when its particle number density calculated explicitly $n_i$, satisfies the Eq. (11):

$$n_i < \beta n^0, \tag{11}$$

where the value of $\beta$ used in this study is 0.85.

*Forces, moment and impulse*

Resultant force ($\mathbf{F}_w$) and moment ($\mathbf{M}_w$) acting on a wall surface ($w$) are calculated by integrating the pressures ($P_i$) of the wall particles $i$ belonging $w$, which is computed by solving the PPE [Eq. (8)]:

$$\mathbf{F}_w = -\sum_{i \in \Omega_w} P_i l_0^2 \mathbf{\eta}_i, \tag{12}$$

$$\mathbf{M}_w = -\sum_{i \in \Omega_w} P_i l_0^2 \mathbf{d}_i \times \mathbf{\eta}_i, \tag{13}$$

where $\mathbf{\eta}_i$ is the normal vector of the wall at wall particle $i$ and $\mathbf{d}_i$ is the perpendicular distance vector from the floor to the line of action of the force at particle $i$.

Fig. 1 illustrates the particle model of a vertical wall ($w$) with its wall particles and dummy particles, as well as the normal vectors, positive outward the wall surface. Also in a similar way, show the wall particles and dummy particles that form the floor.

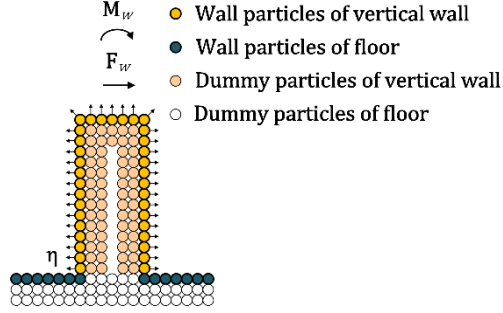

**Fig. 1.** Schematic representations of resultant force ($\mathbf{F}_w$) and moment ($\mathbf{M}_w$) acting on the wall, the normal vectors ($\boldsymbol{\eta}$) and the particles that form the floor.

The impulse (**I**) on the wall is computed considering all the duration of simulation:

$$\mathbf{I} = \sum_{t_0}^{t_F} \mathbf{F}_w \cdot \delta t, \qquad (14)$$

where $t_0$ and $t_F$ is the initial and final time of simulation, respectively.

*Neighbor particle search and high-performance computing*

Searching for neighbor particle is a crucial problem in almost particle-based simulation. Usually, it becomes the bottleneck of the simulation time because it is an O($N^2$) combinatory problem, where *N* is the number of particles, and efficient algorithms are necessary. To increase the efficiency of the computational time cost, the cell index method (Allen and Tildesley 1987) is adopted in the present study.

On the other hand, particle-based methods share the problem of high computational cost associated to large number of particles, necessary to describe realistic applications. Aiming to speed up processing time, a combination of a non-geometric dynamic domain decomposition strategy based on particle renumbering and a distributed parallel sorting algorithm for the particle renumbering proposed by Fernandes et al. (2015) is adopted in the present work for hybrid parallel processing in computer cluster.

*Validation*

Validations showing the effectiveness of the simulation system used in the present study can be found in Belezzi et al. (2013) considering a benchmark case of three-dimensional dam breaking flow hitting a block, the pressure variations inside a tank due to sloshing (Tsukamoto, Cheng, and Nishimoto 2011) and the interaction between free surface flows and elastic plates (Amaro Junior and Cheng 2013).

In the present work, a 3D dam break experiment data provided in Gómez-Gesteira and Dalrymple (2004) and Yeh and Petroff (2003) is considered. The experiment consists of a rectangular tank of 0.61 m wide, 1.60 m long, and 0.75 m high, and a square column with a cross section of 0.12 x 0.12 m, height of 0.75 m, located 0.90 m downstream and aligned with the central axis of the tank. A gate, positioned 0.4 m downstream, holds water to the height of 0.30 m. There is a layer of water 0.01 m high on the tank floor prior to the experiment. The initial geometry is depicted in Fig. 2. Fluid properties and the simulation parameters are given in Table 1 and 2, respectively. The initial value of distance between particles $l_0$ = 2.5 mm and the time step $\Delta t = 0.0004\ s$ were adopted.

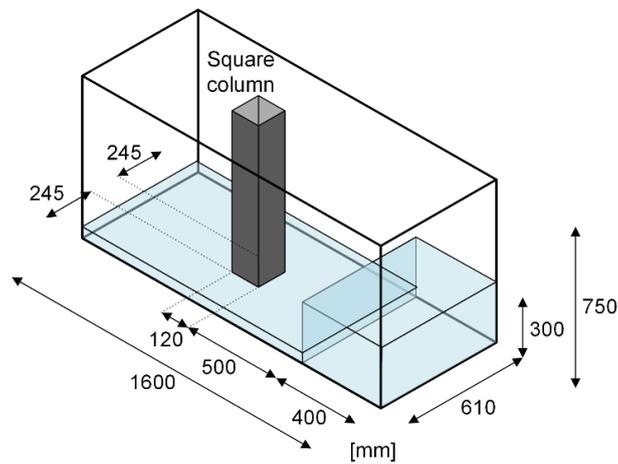

**Fig. 2.** Main dimensions of dam break geometry.

**Table 1.** Fluid properties.

| Property | Value |
|---|---|
| Fluid density ($\rho$) | 1001 kg/m$^3$ |
| Kinematic viscosity ($\nu$) | 10$^{-6}$ m$^2$/s |

**Table 2.** Simulation parameters.

| Parameter | Value |
|---|---|
| Effective radius ($r_e$) | 2.1 $l_0$ |
| Free surface ($\beta$) | 0.85 |
| Relaxation coefficient ($\gamma$) | 0.1 |
| Compressibility factor ($\alpha$) | 10$^{-8}$ ms$^2$/kg |
| Gravitational acceleration ($g$) | 9.81 m/s$^2$ |

The snapshots of the free surface deformation due to the hydrodynamic impact on the square column are presented in Fig. 3. The colors on the free surface of the fluid are related to its velocity field.

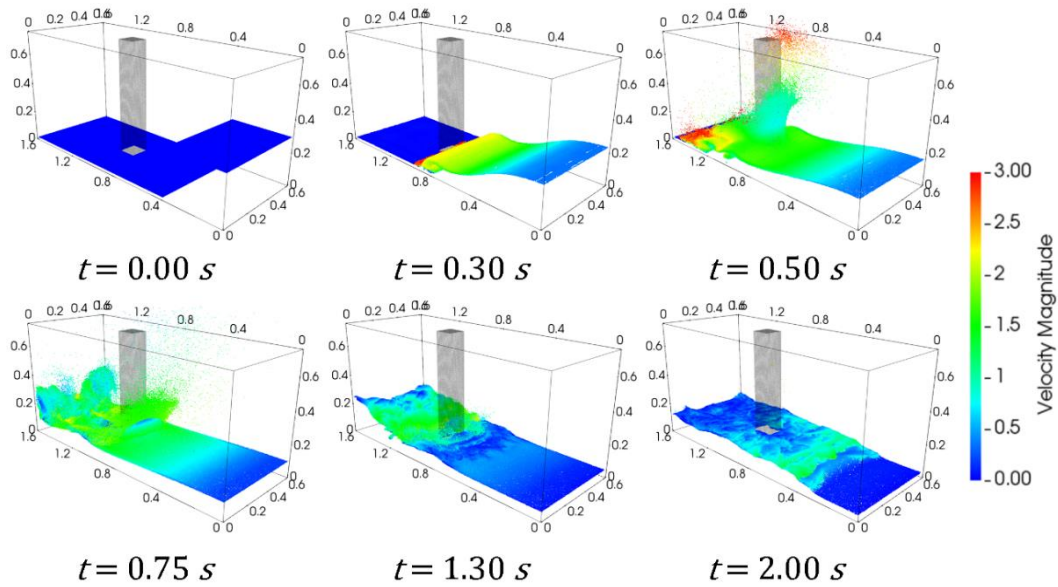

**Fig. 3.** Validation. Snapshots of the free surface deformation.

The experimental and computed resultant forces on the square column are presented in Fig. 4. The computed force is represented by solid line and the experimental data, obtained from four independent experiments, are plotted using circular symbols. The resultant force is calculated by summing the forces that act on both weather and lee sides of the square column. The forces due to the shear stresses on the faces of the column are not taken into account because they are negligible compared to the normal forces on the weather and lee faces. The computed wave front hits against the square column at the instant $t = 0.33\ s$, slightly earlier than the experimental initial impact, which took place at around $t = 0.39\ s$. A reason for this time lag might be lifting of the gate. The velocity that the gate was lifted on the experiment is not available in the references so that, an instantaneous gate opening is considered in the numerical modeling. On the other hand, the value of the computed first peak force on the structure of 34 N, approximately, is in very good agreement with experimental one. Since the impact force is proportional to square of flow velocity ($F \propto \rho A V^2$), the negligible difference between experimentally measured and numerically computed force magnitudes indicates that the tiny time lag has negligible effect on the flow velocity. After the impact, oscillations of the force due to violent free surface deformation and splashing are computed. As the experimental data are not continuous records, this is the main discrepancy between the numerical and experimental results and in overall, the good agreement between the experimental and computed resultant forces show the validity and consistency of the present method to simulates hydrodynamic impact phenomena.

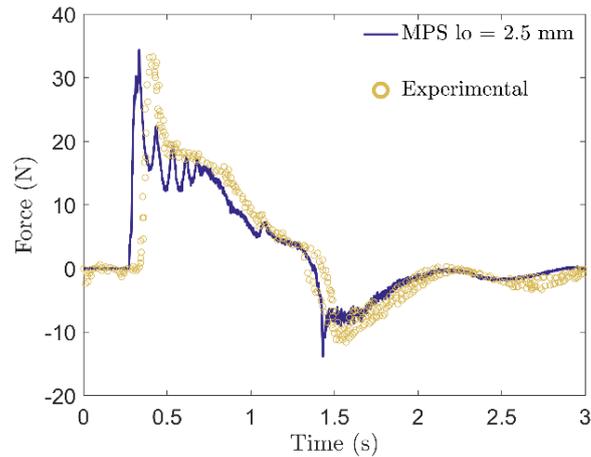

**Fig. 4.** Time histories of resultant force on the square column computed by MPS (solid line) and obtained from four independent experiments (circular symbols).

**Breakwater model**

*Simplified modeling*

According Faltinsen et al. (2001) water on deck can be divided into two main scenarios. One is caused by plunging breakers hitting the deck house or equipment directly. The other scenario is of large relative vertical motions between the ship and the water that creates a wall of water around the deck of which the behavior resembles the one subsequent to a dam breaking. Focusing on the basic issues of the nonlinear hydrodynamic phenomenon in localized regions, in this study the green water flow on deck is modeled considering only dam breaking type green water. The protected installation is modeled herein as a rectangular vertical wall to make easier the interpretation of the hydrodynamic loads. In this way, the incoming wave approximated by a collapsing water column hit first the breakwater, and then the protected vertical wall, as shown in Fig. 5. For the lightweight breakwater, two protection heights and different open-area ratios are used. Both circular and square holes are considered. As well pointed by Buchner (2002), no overtopping is observed in the green water problem so that air entrapment between the structure and the water seems to be less critical. Thus, possible pressure variation due to the air entrapment is neglected in this work.

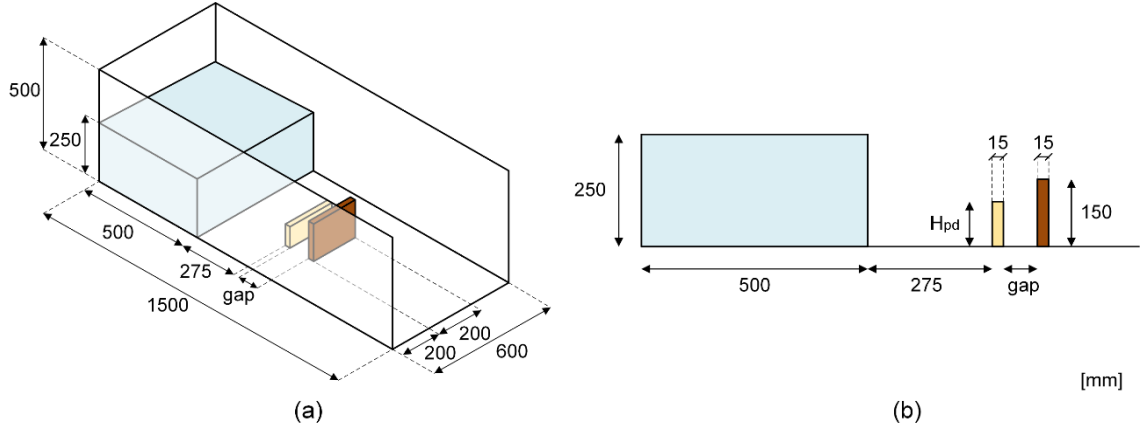

**Fig. 5.** Initial geometry and dimensions of the breakwater model. (a) Breakwater (yellow plate), protected installation (brown wall) and water colunm at initial time. (b) Lateral view.

All numerical simulations are carried out to a model scale of 1:40, in which a wave height of $H_{wave} = 0.111\ m$ (approximately 4.5 m at full scale) are considered, and the water column $H_{dam}$ obtained as (Goda and Miyamoto 1976):

$$H_{dam} = \frac{9}{4} H_{wave} = \frac{9}{4} 0.111 = 0.250\ m. \tag{15}$$

The fluid properties used for all simulations are shown in Table 1.

*Convergence study*

Considering the Courant–Friedrichs–Lewy (CFL) condition (Courant, Friedrichs, and Levy 1967) to avoid numerical instabilities, the approximated maximum flow velocity $|u_{max}| = \sqrt{2gH_{dam}} = 2.22\ m/s$ and adopting limit of Courant number $C = \Delta t |u_{max}|/l_0 < 0.45$, the time step is limited by $\Delta t < 0.2\ l_0$. Since MPS adopts semi-implicit algorithm to solve the incompressible viscous flow, it accepts time steps larger than those for weakly compressible particle methods based on fully explicit schemes, such as SPH or Weakly Compressible Moving Particle Semi-implicit (WCMPS) (Shakibaeinia and Jin 2010), in which the pressure is computed using an equation of state and small time steps are required to reduce the density fluctuation.

To capture the transient hydrodynamic impact phenomena on the walls, all the cases were simulated until 2.5 seconds, when the hydrodynamic loads achieve negligible values. The simulation parameters for all the cases are presented in Table 2.

The initial value of the distance between particles ($l_0$) adopted in the numerical model was defined through a convergence study considering the breakwater with one circular hole of 40 mm diameter ($D$). Four distances of particles were considered: $l_0 = $ 10 mm, 5 mm, 2.5 mm, and 1.25 mm which correspond to resolutions $D/l_0 = $ 4, 8, 16, and 32, respectively. The processing times using the parallelized MPS code (Fernandes et al. 2015) are presented in Table 3.

**Table 3.** Convergence study. Resolution and processing time.

| Distance of particles [mm] | 10.00 | 5.00 | 2.50 | 1.25 |
|---|---|---|---|---|
| Resolution $D/l_0$ | 4 | 8 | 16 | 32 |
| Simulation time [s] | | 2.5 | | |
| Number of particles | 162822 | 944022 | 6159366 | 43804302 |
| Time step [s] | 0.0005 | 0.0005 | 0.0005 | 0.0001 |
| Cluster nodes[a] | 1 | 2 | 4 | 16 |
| Computation time | 0d00h17m | 0d01h23m | 0d09h46m | 7d12h00m |

[a] Cluster with Intel® Xeon® Processor E5 v2 Family, frequency 2.80GHz, 20 cores, and memory 126GB.

Figs. 6(a and b) provide, respectively, the raw force time histories and peak forces on the breakwater for the four resolutions. Similar behaviors were obtained for different resolutions. The computed magnitude of force on the breakwater is lower for the coarse resolutions. For $D/l_0 \geq 16$ the discrepancy becomes smaller, and the oscillation of the computed hydrodynamic loads reduced significantly, indicating more stable computations achieved by the higher resolution models. Moreover, the decrease of the distance between particles leads to the reduction of the discrepancies between peak values that tend to converge to 30 N, approximately, as shown in Fig. 6(b). Confronting the computational time necessary to run high resolution models and tolerant margin of error around 10%, the initial value of the distance between particles $l_0 = $ 2.5 mm ($D/l_0 = $ 16 in this case) is adopted in the following simulations.

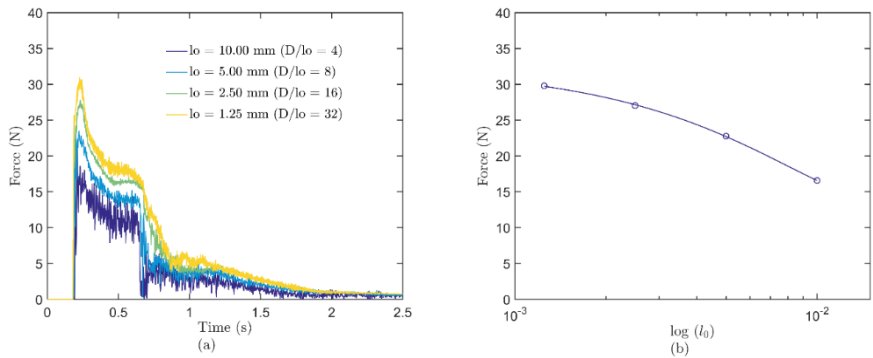

**Fig. 6.** Computed (a) force time histories and (b) peak forces on the breakwater with one circular hole of 40 mm diameter for the four resolutions.

**Results and discussions**

*Cases of study*

The influences of different shapes, number, and the characteristic dimensions of holes, are investigated through a series of numerical simulations. As shown in Fig. 5, a gap of 40 mm between the breakwaters and protected wall was considered initially. Breakwaters with 100 mm height with 1, 2 or 3 circular and square holes were studied first. Since significant volume of fluid may overtops the breakwater and hits the protected wall, the effects of the height of the breakwater were also investigated herein taking into account double height breakwaters with 200 mm height and 2, 4 or 6 circular holes. Also, the influence of hole arrangement was determined using double height breakwaters with new arrangements of 2, 4 and 6 circular holes. Finally, the effects of the spacing between the breakwaters and the protected wall were analyzed using distances of 20, 40, 80 and 160 mm, but considering only 100 mm height breakwaters with 2 circular holes of 30 to 80 mm. The main dimensions of the breakwaters and positions of the holes are shown in Fig. 7. The geometrical dimensions of holes and the open-area ratios of the breakwaters analyzed herein are summarized in Table 4.

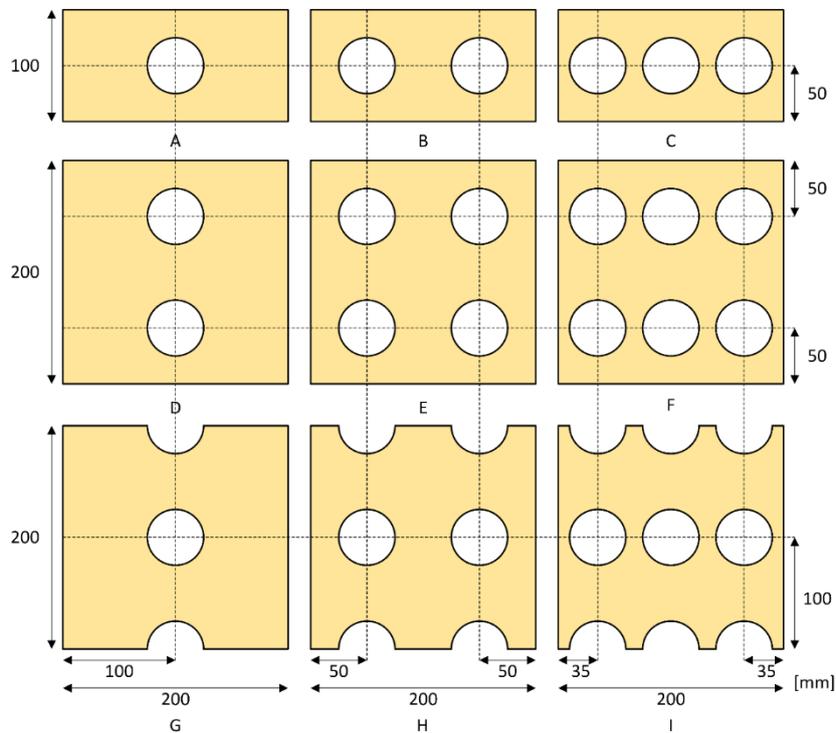

**Fig. 7.** Geometrical dimensions of the 100 mm height breakwaters (models A, B, C), 200 mm height breakwaters (models D, E, F), and 200 mm height breakwaters with new arrangement of holes (models G, H, I).

Table 4. Geometries and dimensions of the holes and open-area ratios of the breakwaters.

| 100 mm height breakwaters | | | | 200 mm height breakwaters | | |
|---|---|---|---|---|---|---|
| Nº of holes | Charac. Dim. (mm) | Open-area ratio (%) | | Nº of holes | Charac. Dim. (mm) | Open-area ratio (%) |
| | | Circular hole | Square hole | | | Circular hole |
| 1 | 30 | 3.53 | 4.50 | 2 | 30 | 3.53 |
| | 40 | 6.28 | 8.00 | | 40 | 6.28 |
| | 50 | 9.82 | 12.50 | | 50 | 9.82 |
| | 60 | 14.14 | 18.00 | | 60 | 14.14 |
| | 70 | 19.24 | 24.50 | | 70 | 19.24 |
| | 80 | 25.13 | 32.00 | | 80 | 25.13 |
| 2 | 30 | 7.07 | 9.00 | 4 | 30 | 7.07 |
| | 40 | 12.57 | 16.00 | | 40 | 12.57 |
| | 50 | 19.63 | 25.00 | | 50 | 19.63 |
| | 60 | 28.27 | 36.00 | | 60 | 28.27 |
| | 70 | 38.48 | 49.00 | | 70 | 38.48 |
| | 80 | 50.27 | 64.00 | | 80 | 50.27 |
| 3 | 30 | 10.60 | 13.50 | 6 | 30 | 10.60 |
| | 40 | 18.85 | 24.00 | | 40 | 18.85 |
| | 50 | 29.45 | 37.50 | | 50 | 29.45 |

*Non-dimensional parameters*

As main output, pressure, force, moment and impulse on the breakwater and pressure, force and impulse on the protected wall are provided. It is important to emphasize that the force on the breakwater is the resultant computed on both weather and lee sides, positive value of the force is oriented to the lee side and positive value of the moment in clockwise direction. The dimensionless values adopted herein are as follows.

The force coefficient ($C_F$) is:

$$C_F = \frac{\|\mathbf{F}\|}{\rho V^2 A}, \tag{16}$$

where $\rho$ is the fluid density, $V$ is the water column velocity approximated by $V = \sqrt{2gH_{dam}}$ and $A$ is the vertical area. Vertical area of the 100 mm height breakwater ($A = 2 \times 10^4 \ mm^2$) is used as reference for all the results to make easy the comparison.

The moment coefficient ($C_M$) is normalized by the hydrodynamic force and the height of the breakwater $H_{pd1} = 100 \ mm$:

$$C_M = \frac{\|\mathbf{M}\|}{\rho V^2 A H_{pd1}}. \qquad (17)$$

The impulse coefficient ($C_I$) is normalized by the hydrodynamic force and the characteristic time $T = \sqrt{H_{dam}/g}$ :

$$C_I = \frac{\|\mathbf{I}\|}{\rho V^2 AT} = \frac{\|\mathbf{I}\|}{2\rho g^{1/2} H_{dam}^{3/2} A}. \qquad (18)$$

Finally, the dimensionless time ($\tau$) is defined as:

$$\tau = t\sqrt{\frac{g}{H_{dam}}}, \qquad (19)$$

where $g = 9.81\ m/s^2$ is the gravitational acceleration.

### *The influence of shape, number and dimension of holes - 100 mm height breakwater*

The performance of a 100 mm breakwater with circular and square holes (models A, B and C shown in Fig. 7 and first column of Table 4) were investigated in this section. The 40 mm gap between the breakwater and the wall was adopted for all cases.

Fig. 8 gives the snapshots of the computed free surface deformation of the collapsing water column hitting the breakwater with 100 mm height and one circular hole of 50 mm diameter. The color scale on the breakwater and protected wall are associated to pressure field and the color scale on the free surface of the fluid is related to velocity magnitude. The fluid hits against the breakwater at the instant $\tau = 1.25$. At the instant $\tau = 3.13$, a large splash is formed, and the protected wall is hit by the fluid that crossed the circular hole or overtopped the breakwater. The collapse of the upward water flow due to gravity occurs around the instant $\tau = 5.01$.

In Fig. 9 the pressure distributions on the weather side of the breakwater and the protected wall are given in the superior and inferior rows, respectively. At the instant $\tau = 1.57$, the maximum pressure is computed for the breakwater and the protected wall. Until $\tau = 3.13$, high impact load occurs only in the central region of the protected wall due to the water jet through the hole of the breakwater. On the other hand, most of the weather side of the breakwater is exposed to relatively high impact pressure. After $\tau =$ 3.13, the pressure on the weather side of the breakwater and part of the protected wall decreases remarkably, and become almost hydrostatic after the instant $\tau = 6.26$.

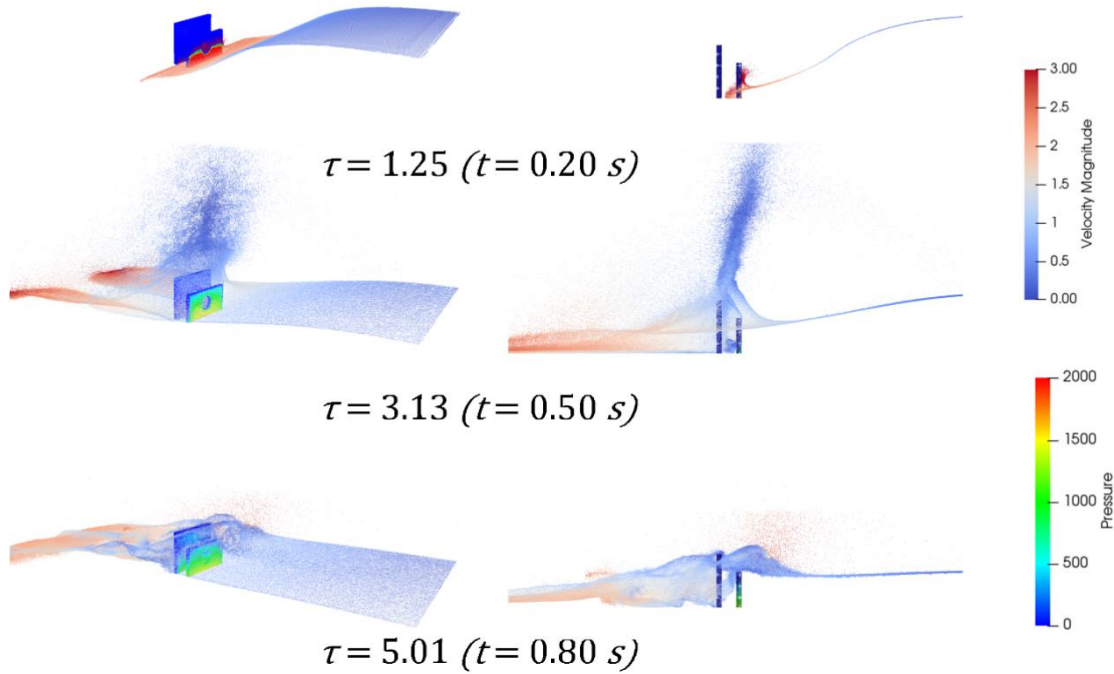

**Fig. 8.** Snapshots of the free surface deformation: 100 mm height breakwater with one circular hole of 50 mm diameter (open-area ratio of 9.82%).

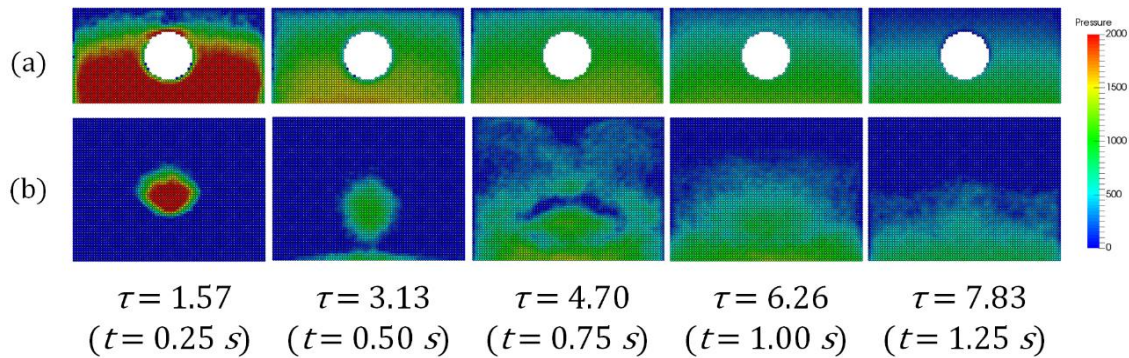

**Fig. 9.** Pressure distribution on the weather side of (a) breakwater and (b) protected wall. Breakwater with 100 mm height and one circular hole of 50 mm diameter (open-area ratio of 9.82%).

Fig. 10(a) and Fig. 11(a) provide the time histories of the computed hydrodynamic impact force on the breakwaters with 100 mm height perforated with one circular hole and one square hole, respectively. Fig. 10(b) and Fig. 11(b) present the time histories of the impact force on the protected wall. Just for reference, the time history of the computed force on the unprotected wall, represented by the dashed line, is also shown.

The incoming water hits the breakwater at $\tau = 1.15$, and hits the protected wall soon later, at $\tau = 1.35$, approximately. The increase of hole dimension leads to lower peak force on the breakwater and higher peak force on vertical walls. Also, drastic force reduction on the breakwater occurs at instant $\tau = 4.10$, caused by the slamming of the back flow on the lee side of the breakwater.

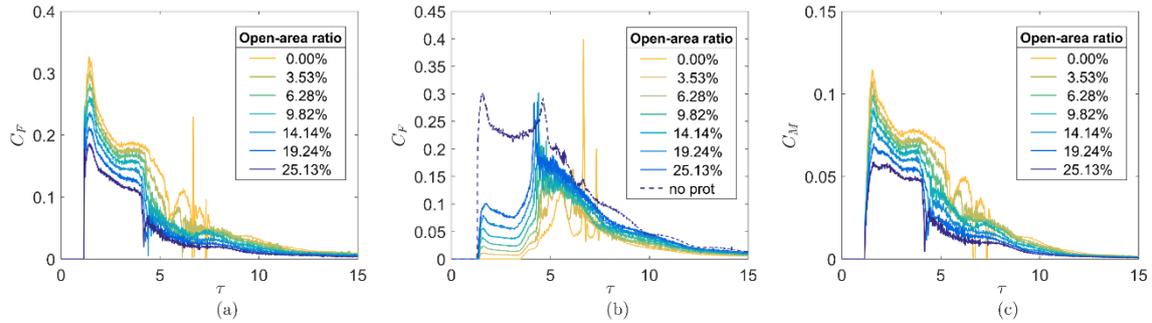

**Fig. 10.** Time histories of the force on (a) breakwaters, (b) protected wall and (c) moment on breakwaters. Breakwaters with 100 mm height and one circular hole (model A of Fig. 7).

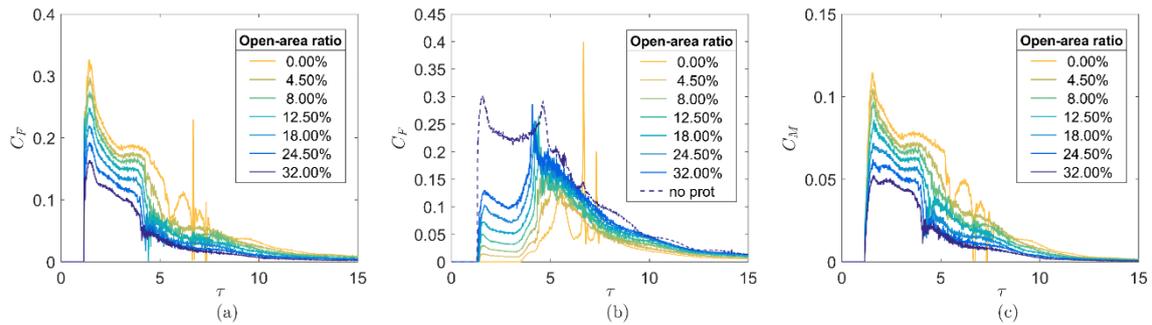

**Fig. 11.** Time histories of the force on (a) breakwaters, (b) protected wall and (c) moment on the breakwaters. Breakwaters with 100 mm height and one square hole (model A of Fig. 7).

The curves of the force on the no perforated breakwater (0% open-area ratio) are quite similar to those of perforated ones, except near the instants $\tau = 6.70$ and $\tau = 7.33$, when high peak forces on the device without perforation occur. At these instants, the fluid that overtopped the breakwater or crossed it through the holes merged inside the small, confined gap between the breakwater and the protected wall. It is important to point out that despite the results suggest these pulses are pressure spikes provoked by the hydrodynamic impact, they could also be caused by inherent numerical oscillations of particle-based methods or numerical modeling neglecting the entrapped air and further investigation through experimental measurements are required. For this case, the force on the protected wall is very low initially. It stays constant for a while and is followed by a much higher impact. After that, the force decreases continuously.

Two high force peaks occur on the unprotected wall. The first and second peaks are due to the incoming wave and the collapse of the upward water flow due to gravity in the region around the protected wall, respectively.

Comparing the results with and without breakwater, the breakwaters reduce the force on the protected wall. However, the reduction of the second peak magnitude at instant $\tau = 4.50$, is relatively small compared to that achieved on the first peak.

Fig. 10(c) and Fig. 11(c) present the computed time histories of the moment on the breakwaters with one circular hole and one square hole, respectively. As the resultant moment is proportional to resultant force, their curves show a similar pattern.

From the comparison between Fig. 10 and Fig. 11, both circular and square holes the curves present similar pattern and the loads are almost independent to hole shape.

The pressure on the weather side of the breakwater with 100 mm height and two circular holes of 80 mm diameter and on the protected wall are presented in Fig. 12. The maximum pressure occurs near $\tau = 1.57$. Between $\tau = 3.13$ and $\tau = 6.26$, a large area of the protected wall is exposed to the water jet through the circular holes. It is interesting to point out the high-pressure zone in the wall center due to the stagnation formed by the interaction of the two jet flows.

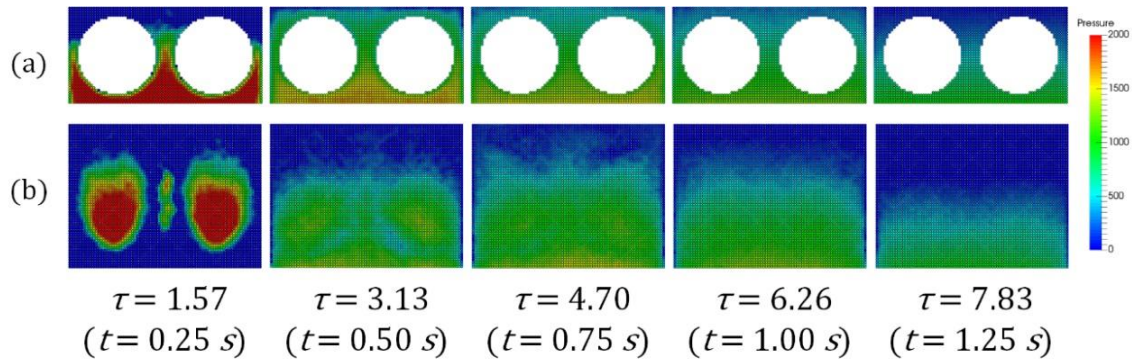

**Fig. 12.** Pressure distribution on the weather side of (a) breakwater and (b) protected wall. Breakwater with 100 mm height and two circular holes of 80 mm diameter (open-area ratio of 50.27%).

Figs 13 and 14 present, respectively, the computed forces and moments regarding the breakwaters with two circular holes and two square holes. Like the breakwater with one hole, the figures show that the increase of hole size leads to lower peak force on the breakwater and higher peak force on the protected wall, and the effects of hole shape are negligible. Considering the same characteristic dimensions of holes, as the open-area ratios of the square holes are larger than the circular ones, higher momentum bypasses the breakwater with square holes, so that the higher initial peak impact load occurs on the wall protected by breakwater with square holes.

The sudden reduction of the hydrodynamic force on the breakwater at $\tau = 4.10$ due to the slamming of the back flow can also be observed. However, for the breakwaters with an open-area ratio above 49%, the hydrodynamic force on the breakwater is small and the associated protected wall suffers high hydrodynamic forces almost nearly to that without protection. Thus, in order to reduce the force on the protected wall, breakwaters with two holes of open-area ratio below 49% should be adopted.

Finally, in the no perforated breakwater (0% open-area ratio), high force peaks due to the merging of fluid and filling up of the gap were also computed from about $\tau = 5.50$ to $\tau = 7.50$, despite the experimental measurements are required to confirm it.

The time histories of the moment on the perforated breakwater given in Figs. 13(c) and 14(c) show that as in the case with one hole, increasing the hole size leads to lower moment. However, when open-area ratio is above 36%, moment higher than the initial peaks were computed at $\tau = 2.50$, approximately. This is because the initial impact force on the breakwater with high open-area ratio is lower, generating the first and small moment peak at $\tau = 1.70$. After that, at about $\tau = 2.50$, the fluid hits the upper region of the breakwater and generates moment higher than the first peak.

Maximum forces on the breakwaters and protected wall as function of open-area ratio are presented in Fig. 14.

According to Fig. 14(a), the maximum forces on the breakwaters decrease monotonically with increase of open-area ratio. Also, the force is almost independent to hole shape, and the relation between the force and open-area ratio is almost linear for the plate with one hole in a wide range of open-area ratio. For plates with two or three holes, this linearity is limited to open-area ratios lower than 13.5%. Above 13.5%, the increase of quantities of hole leads to higher maximum forces.

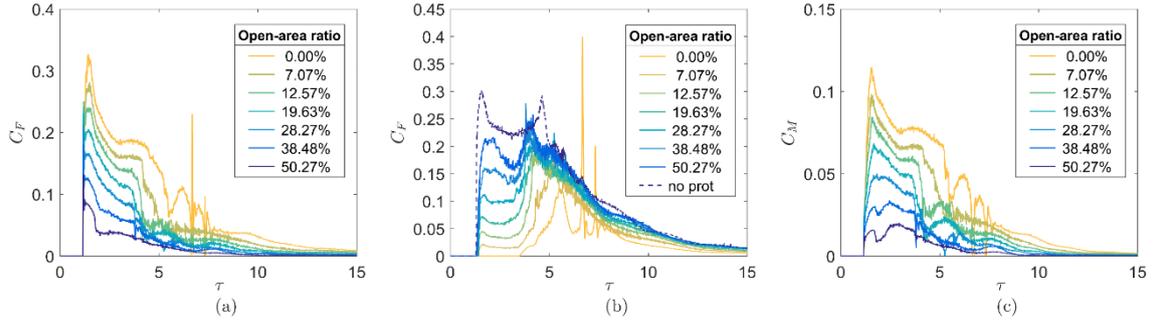

**Fig. 13.** Time histories of the force on (a) breakwaters, (b) protected wall and (c) moment on the breakwaters. Breakwaters with 100 mm height and two circular holes (model B of Fig. 7).

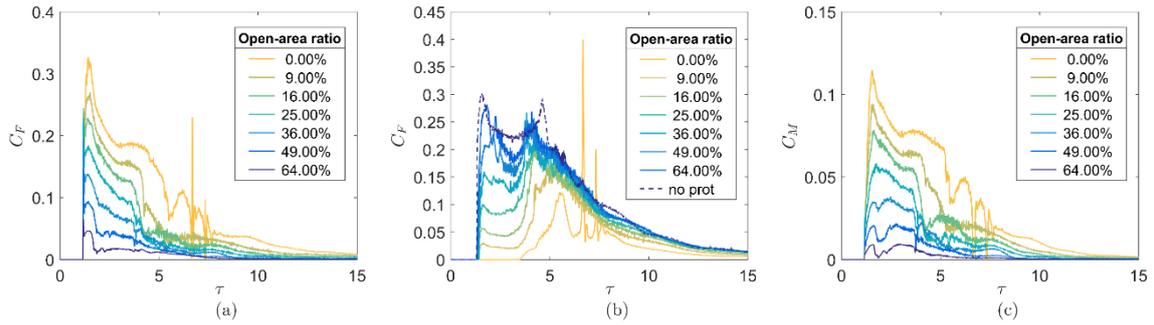

**Fig. 14.** Time histories of the force on (a) breakwaters, (b) protected wall and (c) moment on the breakwaters. Breakwaters with 100 mm height and two square holes (model B of Fig. 7).

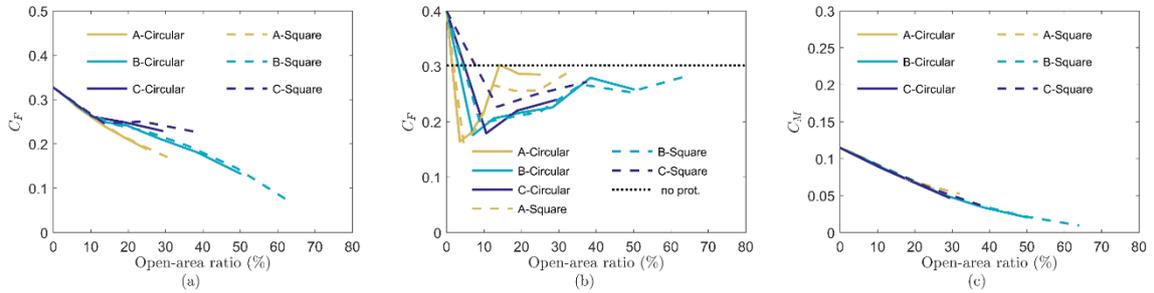

**Fig. 15.** Maximum force coefficient on (a) breakwaters, (b) protected wall and (c) maximum moment on breakwaters. Breakwaters with 100 mm height and circular or square holes (models A, B and C of Fig. 7).

Regarding the protected wall, Fig. 14(b) shows that for the no perforated breakwater (0% open-area ratio), the computed maximum force is higher than without using breakwater. This is associated to very high and short duration force peaks computed for the no perforated breakwater near $\tau = 6.70$ and $\tau = 7.33$, see Figs. Fig. 10(b), 11(b), 13(b) and 14(b), when the flow is entrapped inside the gap between the breakwater and the protected wall. In this way, from the point of view of maximum force on the protected wall, the no perforated plate might be unable to provide desirable protection.

According to Fig. 15(b), the perforated breakwaters with open-area ratio lower than 10% reduce significantly the maximum force, up to 40% lower than the unprotected case, shown by the horizontal solid black line. Also, one-hole breakwaters with open-area ratios over 15% are ineffective because the maximum force on the protected wall becomes almost the same as the unprotected situation.

Fig. 15 (c) shows that the maximum moment decreases with increase in open-area ratio and is independent to shape and quantity of holes.

Fig. 16(a) shows the computed impulse on the breakwater with circular and square holes. The impulse was computed considering 2.5 s ($\tau = 15.66$) of simulation, when the hydrodynamic loads become negligible. Independent of the hole shape, the impulse curves present a slight curvature showing nonlinear relations. The impulse curves of the breakwaters with two or three holes are overlapped and they are lower than those of one hole for open-area ratio above nearly 5%. These results illustrated the complex hydrodynamic owing to flow separation and merging in the perforated plate from single to multiple holes. Another possible reason is the 3-D flow effects due to the relatively low aspect ratio breakwaters considered in the present study.

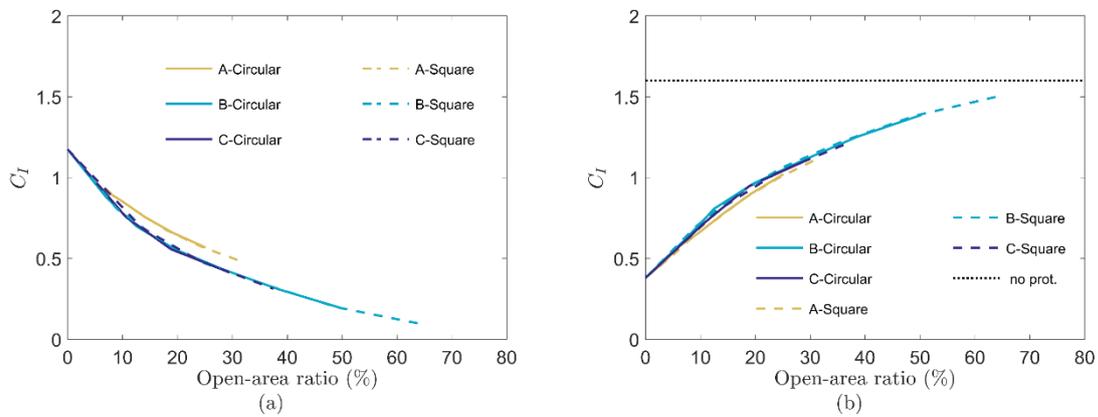

**Fig. 16.** Impulse on the (a) breakwaters and (b) protected wall. Breakwaters with 100 mm height and circular or square holes (models A, B and C of Fig. 7).

The computed impulses on the protected wall are given in Fig. 16(b). Just as a reference, a black solid line that shows the value of the impulse on the unprotected wall (100% open-area ratio), which is $C_I = 1.6$, is also provided. The computed results also show a nonlinear relation between impulse and the open-area ratio. As shown by the impulse on the breakwaters, the curve of one hole is slightly offset from that of two and three holes, while the curves of two and three holes are overlapped. The impulse curves in both hole shapes present the same pattern, where the decrease of open-area ratio leads to an improvement of the wave protection.

*The influence of breakwater height - 200 mm height breakwater*

In the previous cases with 100 mm height breakwaters, fluid overtopping the breakwater and hitting the protected wall may occur. In this section, the performance of a higher breakwater is verified. For this purpose, models D, E and F shown in Fig. 7 and second column of Table 4 were simulated using 40 mm gap between breakwater and the wall.

Fig. 17 gives the snapshots of pressure distribution computed for the breakwater with 200 mm height and two circular holes of 50 mm diameter. The pressures on the protected wall are also shown. The maximum pressure on the 200 mm height breakwater and its protected wall also occurs near $\tau = 1.57$. Until $\tau = 3.13$, the water jet pass only through the lower hole hitting the center of the protected wall. At $\tau = 4.70$, a small portion of the fluid pass through the top hole and hits the central region of the protected wall with much lower intensity. After $\tau = 6.26$, the flow starts to calm down and the hydrodynamic loads on the breakwater and protected wall decreased remarkably. The hydrodynamic pressure only acts on the lower region of the weather side of the breakwater. Also, from the snapshots of $\tau = 4.70$ and $\tau = 6.26$, the region of the protected wall covered by the hydrodynamic load are essentially the same as in the cases with 100 mm height breakwaters, see Fig. 9.

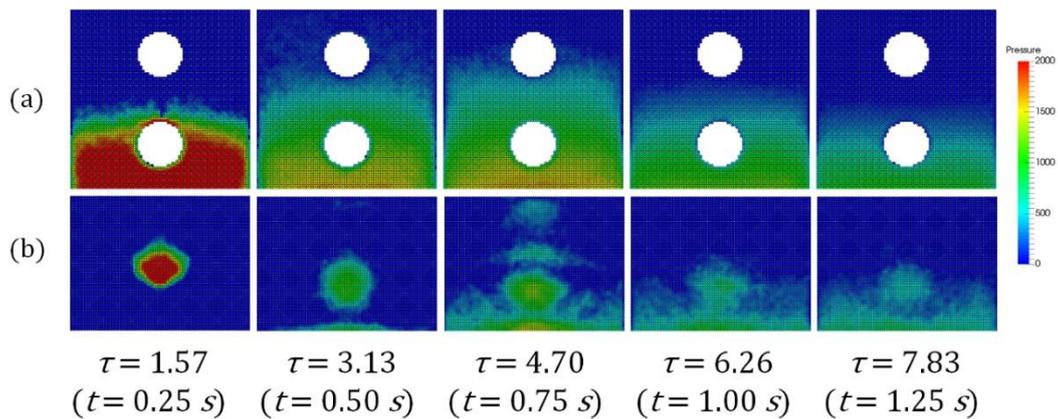

**Fig. 17.** Pressure distribution on the weather side of (a) breakwater and (b) protected wall. Breakwater with 200 mm height and two circular holes of 50 mm diameter (open-area ratio of 9.82%).

Fig. 18(a) provides the time histories of the computed hydrodynamic impact force on the 200 mm height perforated breakwaters with two vertically arranged circular holes (model D of Fig. 7). The forces related to the protected wall are given Fig. 18(b). As in the breakwaters with 100 mm height, the increase of diameter leads to lower peak force on the breakwaters meanwhile the peak force on protected walls are higher. As shown in Fig. 18(a), a second peak force were computed for the breakwaters with 200 mm height at $\tau = 4.30$, when the fluid still hits it. This is different from the breakwaters with 100 mm height, in which the values of force drop significantly at $\tau = 4.30$, see Fig. 10(a). After the second peak force, a drastic reduction of the force on the breakwaters is observed at instant $\tau = 4.40$, approximately, caused by the impact of the back flow on its lee side, after the fluid hits the protected wall.

Compared to the results obtained using 100 mm height breakwaters, see Fig. 10b), increasing the height to 200 mm reduces the force on the protected wall, for all open-area ratios, as depicted in Fig. 18(b). This reduction is remarkable for low open-area ratio breakwaters, and the force on the protected wall is almost null when no perforated (0% open-area ratio) double height breakwater is used. Also, after $\tau = 6.70$ no large magnitude and short duration forces were computed.

Regarding the moment coefficient, very high values were computed between the instants $\tau = 1.50$ and $\tau = 4.20$, approximately, as illustrated in Fig. 18(c). On the other hand, the pattern of the time histories of moment is very different from those of the 100 mm height breakwaters, see Fig. 10(a). For 200 mm breakwaters the maximum $C_M$ is near $\tau = 4.10$, when the splashing water collapses. At this moment, the area of breakwaters weather side covered by the hydrodynamic pressure achieves its maximum, corresponding to lower three forth region of the device, see Fig. 17.

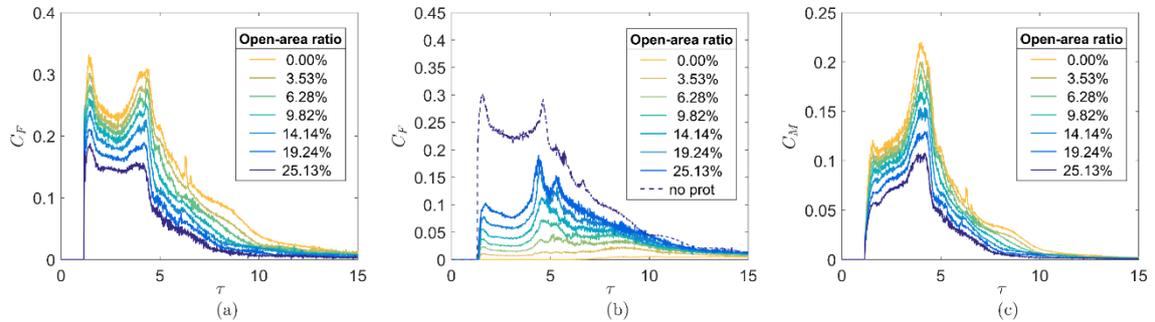

**Fig. 18.** Time histories of the force on (a) breakwaters, (b) protected wall and (c) moment on breakwaters. Breakwaters with 200 mm height and two vertically arranged circular holes (model D of Fig. 7).

Fig. 19 presents the hydrodynamic loads time histories on the 200 mm height four holes breakwaters and on the protected walls. The same tendency of lower peak force on breakwater and higher peak force on the wall with the increase of diameter is recorded.

It is interesting to point out that, as the open-area ratio of the breakwaters increases from 7.07% to 38.48%, the patterns of the force coefficients histories change gradually from that of the 200 mm height plain plate without perforation to that of the highly porous (50.27%) breakwater with 100 mm height. This transition shows clearly that the effects of increasing breakwater height gradually become negligible as breakwater open-area ratio increases. This is because the incoming flow hits essentially the lower part of the device, and in high open-area ratio breakwaters, without significant upward flow deflection in the initial impact, its upper part has no influence on the phenomenon.

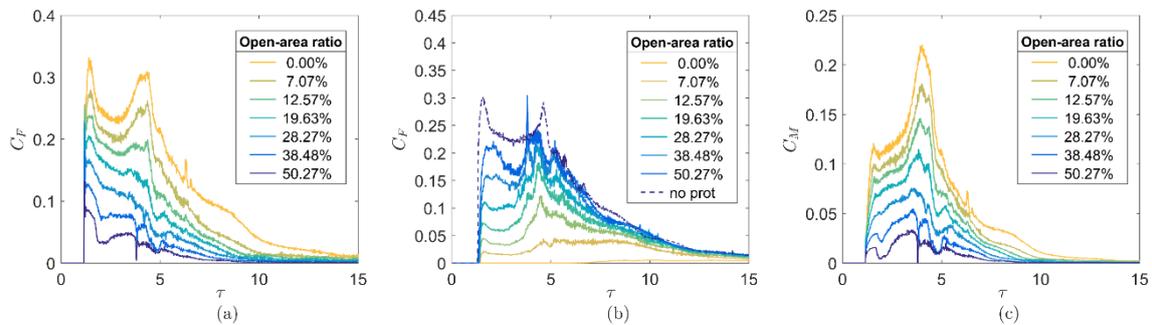

**Fig. 19.** Time histories of the force on (a) breakwaters, (b) protected wall and (c) moment on breakwaters. Breakwaters with 200 mm height and four circular holes (model E of Fig. 7).

Fig. 19(b) shows that the effectiveness of the 200 mm height breakwater is visibly superior to the corresponding 100 mm ones when open-area ratio is lower than 19.63%. The resulting moments on the 200 mm height breakwaters with four circular holes are shown in Fig. 19(c) and the behavior is very similar to those of the 200 mm height breakwaters with two holes, see Fig. 18(c), and equivalent open-area ratio.

*The influence of hole arrangement - 200 mm height breakwater*

The time histories of the computed forces and moments on the 200 mm height breakwaters with four holes in new arrangements (model H of Fig. 7), and respective protected walls are given in Fig. 20. A pattern very similar to the previous one, Fig. 19, is obtained. The difference is that, essentially, semicircular hole in the basement of the breakwaters allows the passage of high velocity flow and impact on the protected wall generating the first peak force followed by a sharp decay between the instants $\tau = 1.00$ and $\tau = 2.00$. Moreover, the computed first peak forces or moments on the breakwaters are slightly higher for the new arrangement.

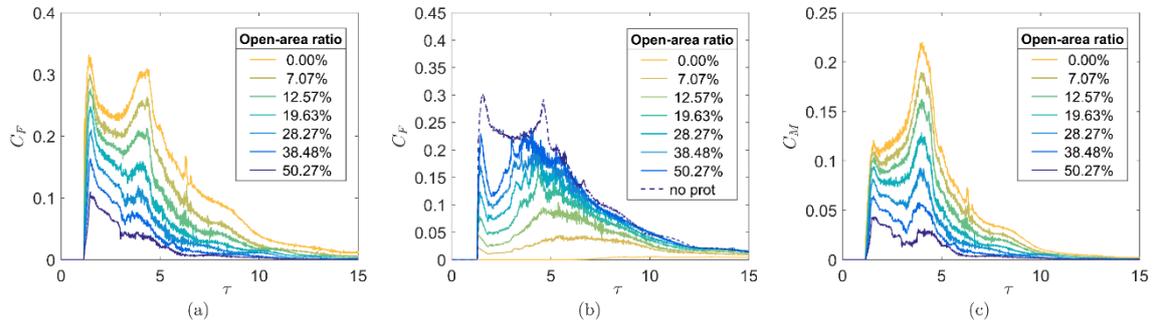

**Fig. 20.** Time histories of the force on (a) breakwaters, (b) protected wall and (c) moment on breakwaters. Breakwaters with 200 mm height and four circular holes in the new arrangement (model H of Fig. 7).

*Maximum forces and moments - 200 mm height breakwater*

Maximum forces on the 200 mm height breakwaters, on the correspondent protected wall and maximum moment on the breakwaters are presented in Fig. 21. As in the cases of 100 mm height breakwater, see Fig. 15, for open-area ratios below 13.5%, the curves of maximum force on the breakwaters are overlapped and decreases linearly with the increase of open-area ratio. For open-area ratios above 13.5%, the increase of the quantity of holes from two to six leads to higher maximum forces on the breakwaters.

On the other hand, the maximum force on the wall protected by the 200 mm breakwaters is directly proportional to the open-area ratio in the range investigated herein. Also, contrary to the cases with 100 mm height breakwaters, the maximum force on the wall protected by 200 mm height no perforated (0% open-area ratio) breakwater is almost zero. This difference might be attributed to the suppression of the flow overtopping the breakwater and thus the entrapment inside the gap. Moreover, compared to 100 mm height breakwaters, see Fig. 15, lower maximum forces on the protected wall are computed for open-area ratios below 30%, showing the improvement achieved by increasing breakwater height. The results also show that open-area ratios above 40% do not offer desirable protection.

The computed maximum moments, shown in (c), decrease with the increase in open-area ratio and the curves of four and six holes are almost overlapped.

In relation to the original arrangement, the breakwaters with new arrangements improve the performance of the breakwater by reducing the maximum forces on the breakwaters and the protected wall, when the open-area ratio is above 25%. The maximum moments of the breakwaters with new arrangement are almost linear and independent of the number of holes for open-area ratios lower than 30%.

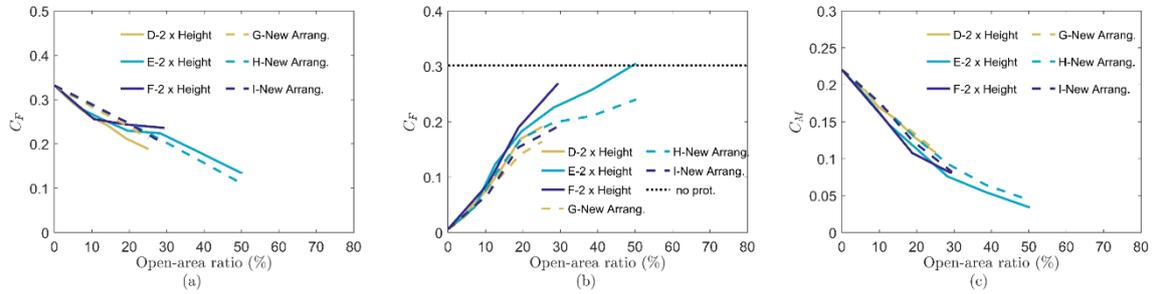

**Fig. 21.** Maximum force coefficient on (a) breakwater, (b) protected wall and (c) maximum moment on breakwater. 200 mm height breakwaters with circular holes and new arrangement of holes (models D, E, F, G, H and I of Fig. 7).

*Impulse on the breakwater and protected wall - 200 mm height breakwater*

The computed impulses on the 200 mm height breakwaters with circular holes are illustrated in (a). Comparing to those with 100 mm height, see Fig. 16, the behaviors are similar but with higher hydrodynamic load for low open-area ratios. The curves show a non-linear relation between the impulse and open-area ratio. Also, the curves of four and six holes are overlapped and slightly lower than the curve of two holes for open-area ratio above 9.82%, showing again that the open-area ratio is the dominant parameter of the phenomenon.

Computed impulses on the wall protected by the 200 mm breakwaters are given in Fig. 22(b). A nonlinear relation between impulse and open-area ratio can be observed for all the cases. As previously observed for the breakwaters, the curves of four and six holes are overlapped and for open-area ratios above 9.82% they are higher than those of two holes. Compared to the walls protected by 100 mm breakwaters, the walls protected by 200 mm breakwaters suffer lower impulse values for open-area ratios below 30%. Thereby based on the cases analyzed herein, the increasing height of breakwaters provides better protection for open-area ratio below 30%.

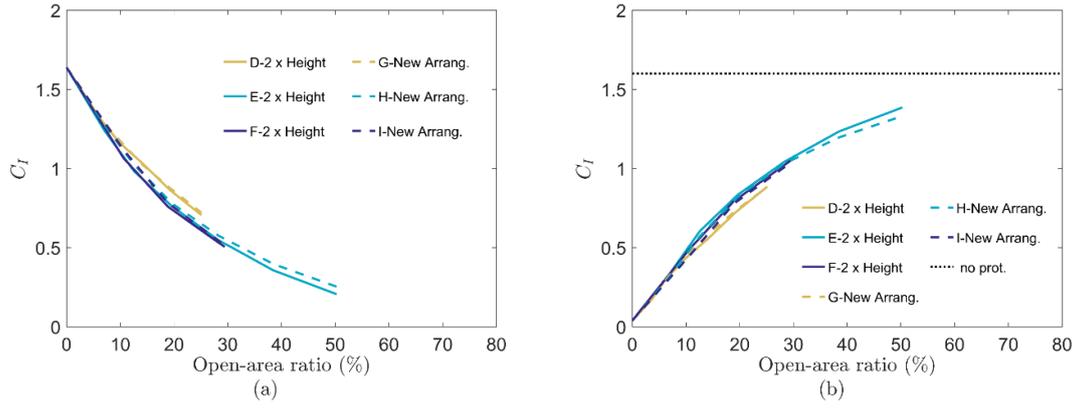

**Fig. 22.** Impulse on the (a) breakwater and (b) protected wall. 200 mm height breakwaters with circular holes and new arrangement of holes (models D, E, F, G, H and I of Fig. 7).

The results of the new arrangement are very close to the original ones, showing that it has no effect on the impulses, which is expected from the momentum conservation.

*The effects of gap between the breakwaters and protected wall*

In the former sections, the gap between the breakwaters and protected wall is 40 mm. In this section, four gap values: $d$ = 20 mm, 40 mm, 80 mm and 160 mm, which correspond to non-dimensional gap $\delta = d/H_{pd1}$ = 0.2, 0.4, 0.8 and 1.6, respectively, are adopted. The positions of the breakwaters are the same as shown in Fig. 5 and the protected wall is displaced to meet the gap. Only 100 mm height breakwaters with two 50 mm diameter circular holes (model B, Fig. 7) are considering in this section.

Fig. 23 shows the computed force time histories on the breakwater and the protected walls. As the position of the breakwater is the same for all cases, the instant and magnitude of the first peak force on the breakwater are independent to the gap and equals to $C_F$ = 0.24. However, the delay of abrupt drop down of the force on the breakwater increases when the gap is increased. This abrupt force reduction is caused by a high hydrodynamic load on the lee side of the breakwater as explained before. On the other hand, the gap greatly influences the impact force on the protected walls. The delay of the first impact on the walls due to enlargement of the gap is clearly shown in Fig. 23(b), as well as a slightly decrease in the magnitude of the initial impact force due to the deceleration of the incoming water.

The patterns of the force time series on the walls change remarkably as the gap varies. In the case with the smallest $\delta$ = 0.2, after the initial impact on the wall, the mean impact force increases steadily. In another extreme, for the largest $\delta$ = 1.6, after the initial contact, the impact force remains almost constant until about $\tau = 4.20$, and then an abrupt rise followed by a smooth decay of the force occurs. Also, smaller peak forces are computed at about $\tau = 6.15$. These patterns result from the forth and back motion of the wave inside the gap. For intermediate $\delta = 0.4$ and $\delta = 0.8$, transition behaviors between two extreme cases were obtained. For $\delta = 0.8$, high magnitude peak force on the protected wall was computed close to $\tau = 5.50$ due to merging of the flows through the holes and that overtopping the breakwater. However, experimental investigations are required to make clear the nature and magnitude of this spike.

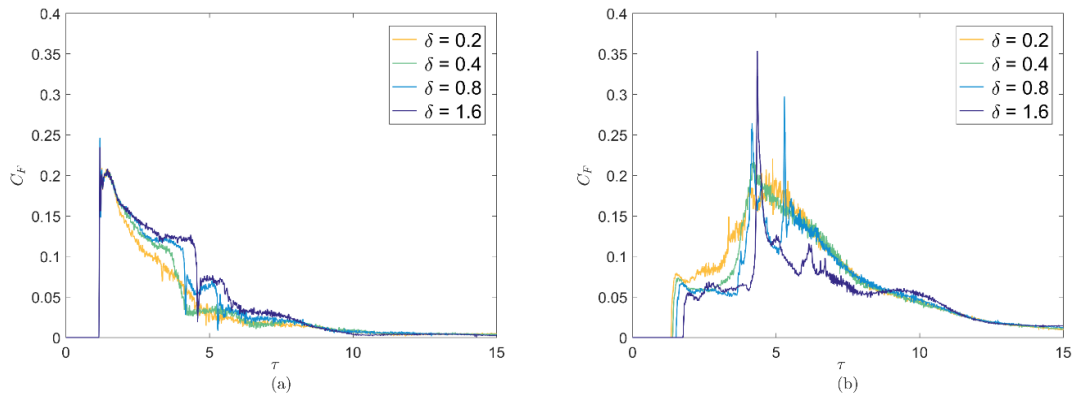

**Fig. 23.** Effects of gap. Force time histories on (a) 100 mm height breakwater with two circular holes of 50 mm diameter (model B of Fig. 7) and (b) protected walls.

Fig. 24 shows the sequences of computed flow and pressure distribution on the longitudinal plane through the hole center. For $\delta = 0.2$, soon after hitting the breakwater, a water jet is formed through the hole and at $\tau = 1.88$, approximately, hits the center of the protected wall. At $\tau = 3.13$, the fluid overtops the breakwater and collides against the upper part of the protected wall resulting in an increasing force and oscillations until $\tau = 5.01$, shown in Fig. 23, as the gap is filled up quickly. On the other hand, in case with $\delta = 1.6$, the water jet that cross the hole hits the upper part of the protected wall at about $\tau = 1.88$. As the jet was directed downward, it hits the base of the vertical wall at $\tau = 3.13$. When the water jet through the hole merged with the water that overtopped the breakwater, high peak force, mainly due the violent plunging jet associated to very complex flow inside the gap, was detected at about $\tau = 4.38$. After that, the flow inside the larger gap results in some lower peak hydrodynamic forces. For intermediate gaps, the computed results show transition patterns.

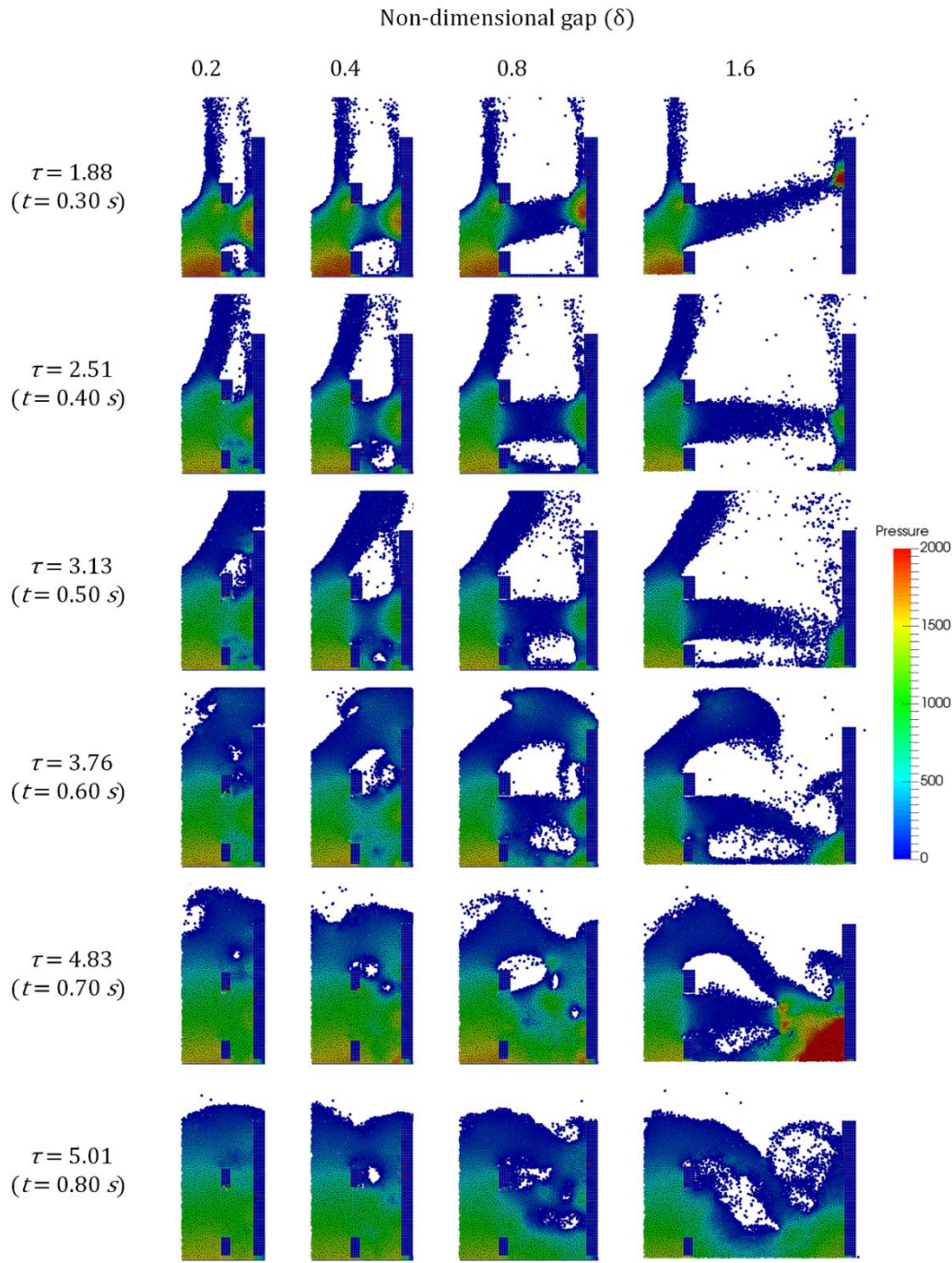

**Fig. 24.** Pressure on the longitudinal plane through the hole center. 100 mm height breakwater with two circular holes of 50 mm diameter (model B of Fig. 7).

Fig. 25 gives the sequences of the computed pressure distribution on the protected wall. As the gap increases, besides the increasing delay of the initial impact on the protected wall, the location where the initial impact occurs is higher. Also, while the region target by the jet flow through the holes of the breakwater is almost fixed and relatively low when the gap is small, as the gap increases, the amplitude of the top-down sweeping motion of the jet flow also increases.

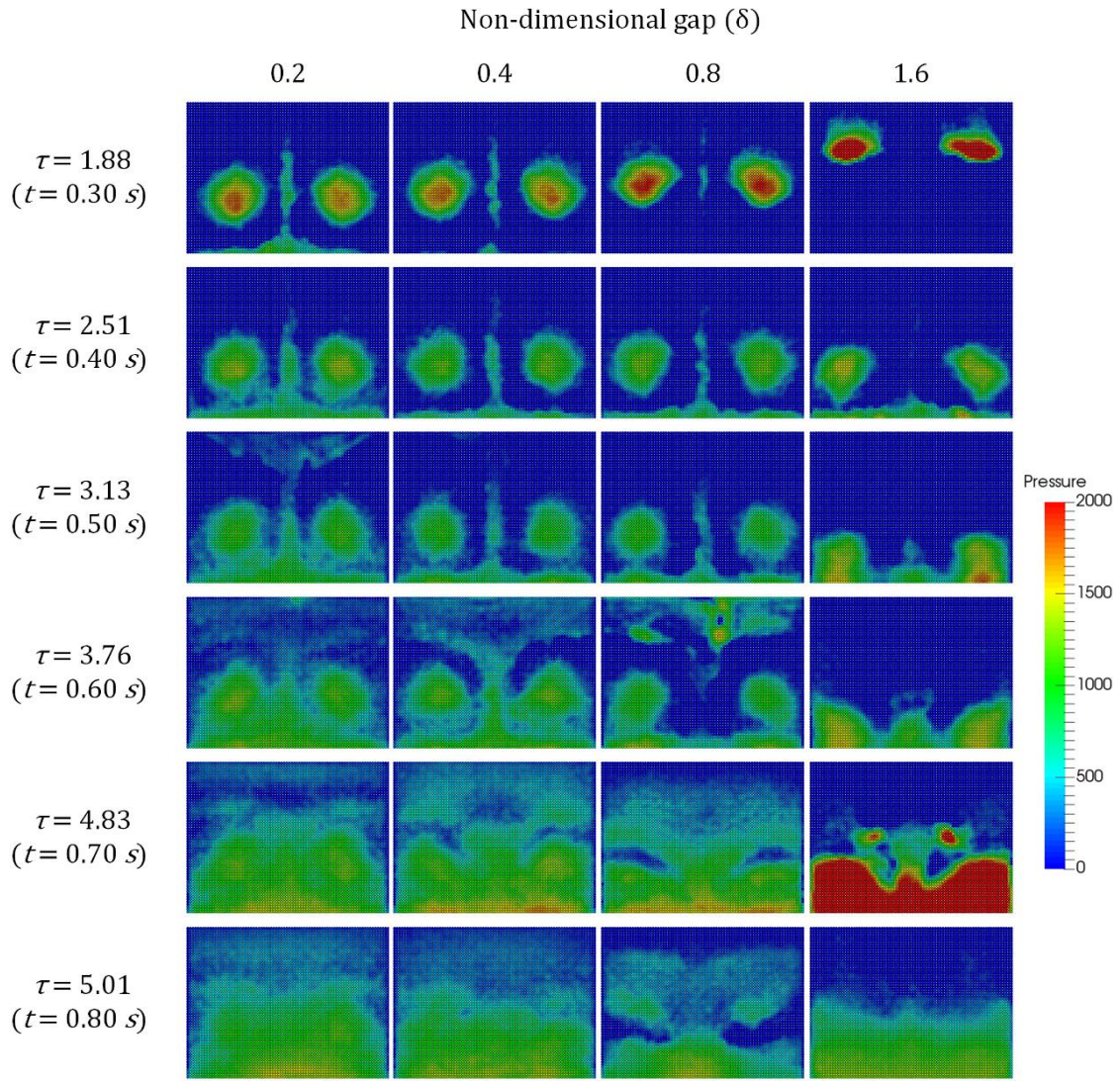

**Fig. 25.** Pressure on the protected wall. 100 mm height breakwater with two circular holes of 50 mm diameter (model B of Fig. 7).

*Applications to the design of perforated breakwater*

Fig. 26 summarizes the loads on the breakwater and protected wall as function of the open-area ratio and the gap. Only 100 mm height breakwaters and two circular holes (model B of Fig. 7) are considered, but as the predominant parameter is open-area ratio, generalization of the results of Fig. 26 is reasonably acceptable. From Fig. 26(a), the impulse on breakwater decreases with increase of open-area ratio, whereas the increase of the gap leads to higher impulse. On the other hand, the increase of open-area ratio or decrease of gap leads to higher impulse on protected wall, as shown in Fig. 26(b).

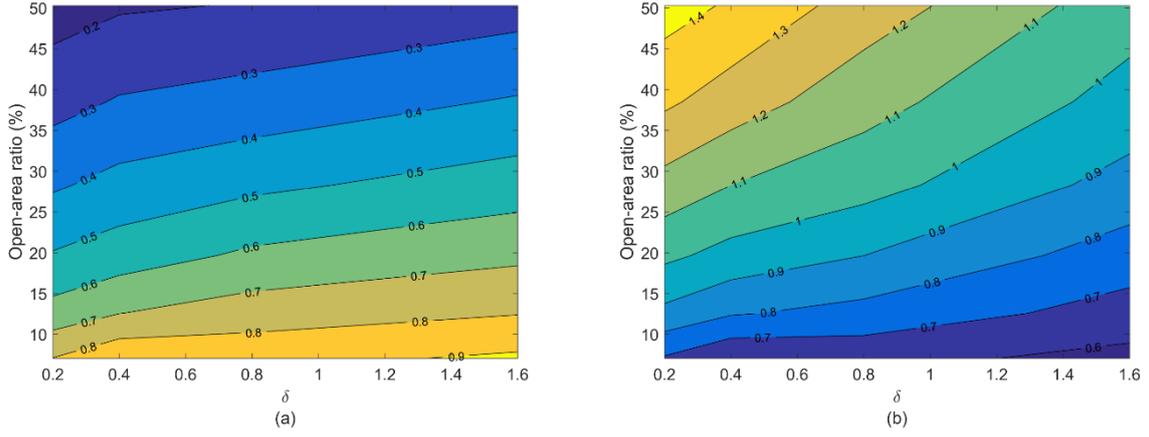

**Fig. 26.** Impulse on (a) 100 mm height breakwater with two circular holes (model B of Fig. 7) and (b) protected wall as a function of the open-area ratio and gap.

In practice, for the optimal design of lightweight perforated breakwater, a goal function might be defined considering its weight, load or other parameters such as the effectiveness of the protection device. The tolerant loads on protected structure and range of gap, which depend on the strength of the structure and the layout on the deck, might be used as restrictions of the design. In this way, as a simple guideline, starting from Fig. 26(b), the corresponding range for open-area ratio of the breakwater might be obtained. Then, through the association of the open-area ratio to the parameter of the goal function, the optimization can be performed with the aid of the Fig. 26(a).

As an example, and for sake of simplicity, a generic goal function $F_I$ that relates the reduction of the load on protected wall, which can be expressed by the coefficients of impulse on protected wall $C_{Iw}$ and on unprotected wall $C_{Io}$, and the reduction of the breakwater weight, which is associated to the open-area ratio $\phi$, can be defined as:

$$F_I(C_I, \phi) = a(C_{Io} - C_{Iw})^b \times \phi^c , \qquad (20)$$

where $a$, $b$ and $c$ are values to be assigned appropriately for a specific design.

Considering $a = b = c = 1.0$, Fig. 27 gives the function $F_I$ for the walls protected by 100 mm height breakwaters with circular and square holes and 200 mm height breakwaters (models A, B C, D, E and F of Fig. 7). In the three cases, optimized open-area ratios are about 30%.

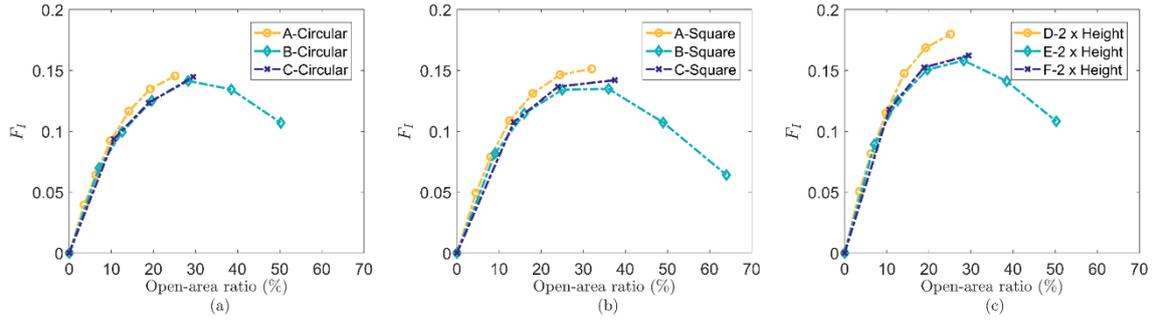

**Fig. 27.** Goal function $F_I$ considering 100 mm height breakwaters with (a) circular holes, (b) square holes and (c) 200 mm height breakwater with circular holes (models A, B C, D, E and F of Fig. 7).

**Final considerations**

In the present study, the influences of geometrical parameters of perforated breakwaters on their performance, as well as the effects of the gap were investigated by a particle-based simulation method and modeling the incoming wave as a collapsing water column subjected to gravity.

The results show that the general patterns of the hydrodynamic force and moment time histories are similar regardless of shapes, number, and arrangement of holes. The primary parameter affecting the performance of perforated breakwater is open-area ratio.

The peak force, maximum moment and impulse on the breakwater are inversely proportional to the open-area ratio, while the hydrodynamic loads on the protected device are proportional, despite the relations not being linear. In order to reduce the force on the protected wall, an open-area ratio below 49% should be adopted.

Increasing height of the breakwater is effective to mitigate the loads on the protected wall only for open-area ratios below 30%. This is because the incoming flow hits essentially on the lower part of the device and as open-area ratio increases, upward deflection decreases, and upper part of the device becomes ineffective. In this way, with increasing open-area ratio, the patterns of the time histories of the hydrodynamic loads change gradually to that of 100 mm height.

Concerning the gap, larger gap increases the delay of the initial impact on the protected wall and the top-down sweep motion of the jet flow through the holes. Also, the increase of the gap leads to slightly higher impulse on the breakwater, while decreases the impulse on the protected wall.

Finally, the results also provide explicitly the relations between the hydrodynamic loads on the breakwater and the protected wall, which can be adopted as a basis for the optimal design of breakwater considering the tolerant hydrodynamic impact loads and geometrical restrictions on deck. Also, they can be extended to coastal protection devices by considering sequence of incoming waves.

**Acknowledgments**

This work has financial support from Coordenação de Aperfeiçoamento de Pessoal de Nível Superior - Brasil (CAPES) - Finance Code 001 and FAPESP Proc. No. 2015/10287-0. The authors are grateful to Petrobras for financial support on the development of the MPS/TPN-USP simulation system based on MPS method. The authors would like to express sincere thanks to D.Sc. Daniel Fonseca de Carvalho e Silva for valuable discussions.